\documentclass[prd,aps,nofootinbib,superscriptaddress]{revtex4}
\usepackage{epsfig}
\usepackage{amsmath}
\usepackage{amsfonts}
\usepackage{amssymb}
\usepackage{color}
\usepackage{graphicx}
\usepackage{graphics}
\usepackage{hyperref}
\usepackage{epstopdf}

\begin{document}
\title{Radiative Neutrino Mass Model at the $e^{-}e^{+}$ Linear Collider}
\author{Amine Ahriche}
\email{aahriche@ictp.it}
\affiliation{Department of Physics, University of Jijel, PB 98 Ouled Aissa, DZ-18000 Jijel, Algeria.}
\affiliation{The Abdus Salam International Centre for Theoretical Physics, Strada Costiera
11, I-34014, Trieste, Italy.}
\author{Salah Nasri}
\email{snasri@uaeu.ac.ae}
\affiliation{Physics Department, UAE University, POB 17551, Al Ain, United Arab Emirates.}
\affiliation{Laboratoire de Physique Theorique, ES-SENIA University, DZ-31000 Oran, Algeria.}
\author{Rachik Soualah}
\email{rsoualah@cern.ch}
\affiliation{INFN Sezione di Trieste, Gruppo Collegato di Udine, University of Udine, via
delle Scienze, 208, I-33100 Udine, Italy.}

\begin{abstract}
We study the phenomenology of a Standard Model extension with two charged
singlet scalars and three right handed neutrinos at an electron-positron
collider. In this model, the neutrino mass is generated radiatively at
three-loop, the lightest right handed neutrino is a good dark matter
candidate, and the electroweak phase transition strongly first order as
required for baryogenesis. We focus on the process $e^{+}+e^{-}\rightarrow
e^{-}\mu^{+}+E_{miss}$, where the model contains new lepton flavor violating
interactions that contribute to the missing energy. We investigate the
feasibility of detecting this process at future $e^{-}e^{+}$ linear colliders
at different center of mass energies: $E_{CM}=250,~350,~500~\mathrm{GeV}$, and
$1$ $\mathrm{TeV}$.

\end{abstract}
\maketitle

\section{Introduction}

The Standard Model (SM) of elementary particle physics is very successful in
explaining physics around the electroweak scale. However, despite this several
questions remain to be answered such as neutrino masses and mixing \cite{nu},
the nature of the dark matter (DM) \cite{DM}, and the origin of the baryon
asymmetry of the Universe \cite{BAU}. None of these issues is successfully
explained within the SM. Therefore, various extensions beyond the SM have been
proposed to address these problems.

In Ref. \cite{KNT}, Krauss, Trodden, and one of the authors in this paper,
proposed an extension of the SM with two charged $SU(2)_{L}$ singlet scalars
and one right handed (RH) neutrino field $N_{1}$, where a $\mathbb{Z}_{2}$
symmetry was imposed at the Lagrangian level in order to forbid the Dirac
neutrino mass terms. After the breaking of the electroweak symmetry, neutrino
masses are generated at three-loop, which makes their masses naturally small
due to the high loop suppression. Moreover, the field $N_{1}$ is odd under
$\mathbb{Z}_{2}$ symmetry, and thus it is guaranteed to be stable, which makes
it a good candidate for DM. Ref. \cite{Cheung} studied the phenomenological
implication of this model with two RH neutrinos, instead of just one. In
\cite{hna}, it was shown that in order to fit the neutrino oscillation data
and be consistent with different recent experimental constraints such as
lepton flavor violation (LFV), one needs to have three RH neutrinos. Somewhat
similar classes of three-loop neutrino mass models have also been studied in
\cite{AKS, AMN}.

In this work, we consider the feasibility of testing this radiative model at
the next-generation electron-positron colliders
\cite{ILC1,ILC2,ILC3,CLIC,tlep}. The International Linear Collider (ILC),
being designed for operation at several $e^{-}e^{+}$ collision energies, will
be a great opportunity to anticipate detailed physics studies of our model.
Among the different processes which can be studied at the ILC, we will focus
here on the process $e^{-}e^{+}\rightarrow e^{-}\mu^{+}+E_{miss}$ within the
allowed kinematic regions of the machine. Further dedicated studies that probe
different final states are in preparation for future works \cite{NEXT}.
Accordingly, our signal will consist of electron, anti-muon and missing
energy. In the SM, the missing energy is coming just from one source
$E_{miss}^{(SM)}\equiv\bar{\nu}_{e}\nu_{\mu}$, whereas in our model, there are
twelve different processes that give rise to $E_{miss}$ in the final state:
six with SM left handed (LH) neutrinos and six contributions with heavy RH
Majorana neutrinos. Moreover, the background process in our model gets
modified through extra additional channels. Thus, we look for the excess in
the number of events from the process $e^{-}e^{+}\rightarrow e^{-}\mu
^{+}+E_{miss}$ in this model over the contribution from the SM and then
identify whether the missing energy is produced from LH or RH neutrinos.
Similar effects have been investigated in other class of models \cite{kan,tan}.

This paper is organized as follows: in section II, we introduce the three-loop
radiative model for neutrino masses, then we discuss its detectability at
linear collider in section III. In section IV, we present and discuss the
simulation results. We give our conclusion in section V.

\section{The Model}

\begin{figure}[h]
\begin{centering}
\includegraphics[width=5cm,height=2.2cm]{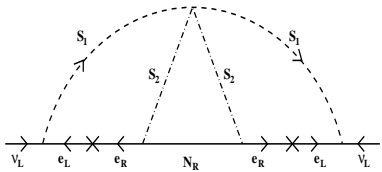}
\par\end{centering}
\caption{The three-loop diagram that generates the neutrino mass matrix
elements.}%
\label{diag}%
\end{figure}

The model that we will study is an extension of the SM with three RH
neutrinos, $N_{i}$, and two electrically charged scalars, $S_{1}$ and $S_{2}$,
that are singlet under the $SU(2)_{L}$ gauge group in addition to a discrete
$\mathbb{Z}_{2}$ symmetry, under which $\{S_{2},N_{i}\}\rightarrow
\{-S_{2},-N_{i}\}$ and all other fields are even. The Lagrangian reads
\cite{hna}%
\begin{equation}
\mathcal{L}=\mathcal{L}_{SM}+\{f_{\alpha\beta}L_{\alpha}^{T}Ci\tau_{2}%
L_{\beta}S_{1}^{+}+g_{i\alpha}N_{i}^{C}\ell_{\alpha R}S_{2}^{+}+\tfrac{1}%
{2}m_{N_{i}}N_{i}^{C}N_{i}+h.c\}-V(\Phi,S_{1},S_{2}),\label{L}%
\end{equation}
where $L_{\alpha}$ is the LH lepton doublet, $f_{\alpha\beta}$ are the Yukawa
couplings which are antisymmetric in the generation indices $\alpha$ and
$\beta$, $m_{N_{i}}$\ are the Majorana RH neutrino masses, $C$ denotes the
charge conjugation operator, and $V(\Phi,S_{1},S_{2})$ is the tree-level
scalar potential which is given by%
\begin{align}
V(\Phi,S_{1,2}) &  =\lambda\left(  \left\vert \Phi\right\vert ^{2}\right)
^{2}-\mu^{2}\left\vert \Phi\right\vert ^{2}+m_{1}^{2}S_{1}^{\ast}S_{1}%
+m_{2}^{2}S_{2}^{\ast}S_{2}+\lambda_{1}S_{1}^{\ast}S_{1}\left\vert
\Phi\right\vert ^{2}+\lambda_{2}S_{2}^{\ast}S_{2}\left\vert \Phi\right\vert
^{2}\nonumber\\
&  +\frac{\eta_{1}}{2}\left(  S_{1}^{\ast}S_{1}\right)  ^{2}+\frac{\eta_{2}%
}{2}\left(  S_{2}^{\ast}S_{2}\right)  ^{2}+\eta_{12}S_{1}^{\ast}S_{1}%
S_{2}^{\ast}S_{2}+\left\{  \lambda_{s}S_{1}S_{1}S_{2}^{\ast}S_{2}^{\ast
}+h.c\right\}  ,\label{V}%
\end{align}
with $\Phi$ denoting the SM Higgs doublet. It has been shown that this model
has the following features \cite{hna}:

$\bullet$ Small non-zero neutrino masses are generated radiatively at
three-loop as shown in Fig. \ref{diag}, which fits the neutrino oscillation
data and without being in conflict with several experiential constraints such
as the bounds on lepton flavor violating processes, the muon anomalous
magnetic moment and the neutrino-less double beta decay.

$\bullet$ Has a DM candidate ($N_{1}$) with a relic density in agreement with
the observation for masses around the electroweak scale.

$\bullet$ Gives rise to a strong first order phase transition that is required
for a successful baryogenesis without being in conflict with the recent Higgs
mass measurements provided by the ATLAS \cite{ATLAS} and CMS \cite{CMS} Collaborations.

$\bullet$ Possible enhancement in the Higgs decay channel $h\rightarrow
\gamma\gamma$, while the channel $h\rightarrow\gamma Z$ gets a small
suppression within 5\% according to the SM.

$\bullet$ Significant large enhancement on the triple Higgs coupling due to
the extra contributions.

\section{ Phenomenology at Linear Colliders}

Neutrinos, which manifest inside the detector as missing energy, are produced
due to the new lepton flavor violating interactions given by the $f$ and $g$
terms in (\ref{L}). At an $e^{-}e^{+}$ linear collider such as the ILC, they
can be directly pair produced in association with a single (or multiple)
photon(s), or a pair of charged leptons\footnote{At the LHC, the LH (RH)
neutrino can be produced via the decay of the charged scalar $S_{1}$ ($S_{2}%
$). However, the production rate of $S_{1,2}$ is expected to be small at the
LHC.}. In addition, both light ($\nu_{i}$) and heavy ($N_{i}$) neutrinos can
be generally produced at linear colliders according to their production cross
section. However, by using polarized beams, one can reduce/increase the
production rate of one type of chiral particles as compared to the other ones.
For instance, if one uses LH polarized electron beam, the heavy RH Majorana
neutrinos production rate gets suppressed and vice-versa. First, we will
concentrate on the process $e^{-}e^{+}\rightarrow e^{-}\mu^{+}+E_{miss}$
considering unpolarized beams for different center of mass (CM) energies,
$E_{CM}$, that can be accessible at the ILC. Second, we generalize our
analysis by allowing for the possibility of tuning the beam polarization at
the linear colliders.

As a benchmark, we consider the following set of the model parameter values:
\begin{align}
f_{e\mu} &  =-(4.97+i1.41)\times10^{-2},~f_{e\tau}=0.106+i0.0859,~f_{\mu\tau
}=(3.04-i4.72)\times10^{-6},\nonumber\\
g_{i\alpha} &  =10^{-2}\times\left(
\begin{array}
[c]{ccc}%
0.2249+i0.3252 & 0.0053+i0.7789 & 0.4709+i1.47\\
1.099+i1.511 & -1.365-i1.003 & 0.6532-i0.1845\\
122.1+i178.4 & -0.6398-i0.6656 & -10.56+i68.56
\end{array}
\right)  ,\nonumber\\
m_{N_{i}} &  =\{162.2\mathrm{~GeV},\ 182.1\mathrm{~GeV},~209.8\mathrm{~GeV}%
\},~m_{S_{i}}=\{914.2\mathrm{~GeV},~239.7\mathrm{~GeV}\},\label{bA}%
\end{align}
which has all the features listed in the previous section. For this benchmark,
the only kinematically allowed decay modes of $N_{2,3}$ are three-body decays
and therefore their total decay widths are small. This means they might decay
outside the detector, and therefore their signatures are similar to the of
$N_{1}$. Detailed studies about the production mechanisms and the decay modes
in $e^{+}e^{-}$ collisions, of new heavy fermions and neutrinos has been
performed in \cite{Dj1}. Furthermore, the analysis of various signals and
backgrounds of new heavy fermions predicted by the SM extensions can be found
in \cite{Dj2}.

In our model, the missing energy in the process $e^{-}e^{+}\rightarrow
e^{-}\mu^{+}+E_{miss}$ corresponds to any state in the set $\mathcal{E}%
_{miss}\subset\{\nu_{\mu}\bar{\nu}_{e}$, $\nu_{e}\bar{\nu}_{\tau}$, $\nu
_{\tau}\bar{\nu}_{e}$, $\nu_{\mu}\bar{\nu}_{\mu}$, $\nu_{\tau}\bar{\nu}_{\mu}
$, $\nu_{\tau}\bar{\nu}_{\tau}$, $N_{i}N_{k};~i,k=1,2,3\}$. The total expected
cross section of the processes $e^{-}e^{+}\rightarrow e^{-}\mu^{+}+E_{miss}$
is represented by $\sigma^{EX}$, while $\sigma(\mathcal{E}_{miss})$ denotes
the cross section of different subprocesses.

The background comprises two leptons (electron and anti-muon) plus missing
energy $E_{miss}^{(SM)}\equiv\bar{\nu}_{e}\nu_{\mu}$.\textbf{\ }In our model,
the subprocess $\mathcal{E}_{miss}\equiv\nu_{\mu}\bar{\nu}_{e}$ has 22
diagrams mediated by $S_{1}$ in addition to the 18 diagrams that exist in the
SM. Then, the number of signal events is the difference between the
contributions from the 12 subprocesses mentioned above and the SM background.
Hence, our goal will be to study the feasibility of detecting any possible
excess of events in our model compared to the SM predictions at the ILC
\cite{ILC1,ILC2,ILC3,ILC4} for different beam energies. Our analysis also
applies to the other future leptonic linear colliders, such as Compact Linear
Collider (CLIC) \cite{CLIC}; and the Triple-Large Electron-Positron Collider
(TLEP) \cite{tlep}.

For small beam energies (such as 250 or 350 \textrm{GeV}), the RH neutrinos
cannot be produced due to the kinematical constraints; and when there are any
observed events from the defined signal, it is due to the light LH neutrinos.
However, at higher energies up to around 1 \textrm{TeV}, one expects the heavy
RH neutrinos ($N_{i}$) to be pair produced and therefore contribute
significantly to the total cross section of the process $e^{-}e^{+}\rightarrow
e^{-}\mu^{+}+E_{miss}$. In this case, the energy of the muon and/or electron
is expected to be smaller than the case where the missing energy is LH neutrinos.

\begin{figure}[t]
\begin{centering}
\includegraphics[width=8cm,height=5.5cm]{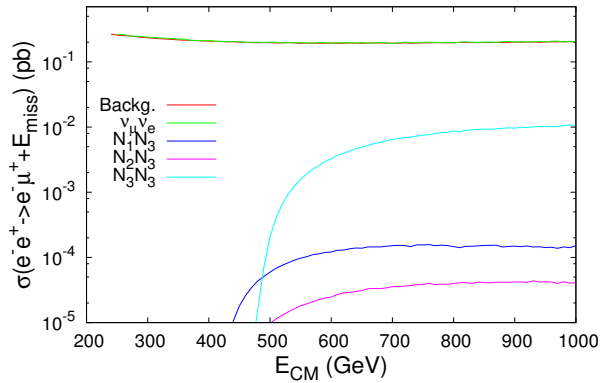}
\par\end{centering}
\caption{The cross section of different contributions $\sigma(\mathcal{E}%
_{miss})$ versus the center of mass energy. The difference between the red and
green curves that represent the background and the subprocess $\mathcal{E}%
_{miss}\equiv\bar{\nu}_{e}\nu_{\mu}$ can not be seen due to the considered
range. The curves $N_{i}N_{3}$ are the only non-negligible contributions that
can be presented for the benchmark in (\ref{bA}).}%
\label{Svss}%
\end{figure}

Fig. \ref{Svss} shows the cross section for each subprocess versus the CM
energies $E_{CM}=250\sim1000$ \textrm{GeV} using unpolarized beams for the
considered model parameters given in (\ref{bA}). Among the different
contributions to the total missing energy, the cross section for the final
states $\mathcal{E}_{miss}\subset\{\nu_{e}\bar{\nu}_{\tau}$, $\nu_{\tau}%
\bar{\nu}_{e}$, $\nu_{\mu}\bar{\nu}_{\mu}$, $\nu_{\tau}\bar{\nu}_{\mu}$,
$\nu_{\tau}\bar{\nu}_{\tau}$, $N_{1}N_{1}$, $N_{1}N_{2}$, $N_{2}N_{2}\}$ are
found to be negligible. In Fig. \ref{Svss}, we show the plot of $\sigma
(\mathcal{E}_{miss}\equiv N_{i}N_{3})$, with $i=1,2,3$, versus the CM energy,
which shows that $\sigma(\mathcal{E}_{miss}\equiv N_{3}N_{3})$ is much larger
than both $\sigma(\mathcal{E}_{miss}\equiv N_{1}N_{3})$ and $\sigma
(\mathcal{E}_{miss}\equiv N_{2}N_{3})$. Then at low CM energies, the signal in
this model, comes only from the subprocess $e^{+}+e^{-}\rightarrow e^{-}%
\mu^{+}\nu_{\mu}\bar{\nu}_{e}$, i.e., the diagrams that are mediated by the
charged scalar $S_{1}$. At higher values of $E_{CM}$, there could
be\textbf{\ }additional contributions from the subprocesses $e^{+}%
+e^{-}\rightarrow e^{-}\mu^{+}N_{i}N_{3}$, which are mediated by the charged
scalar $S_{2}$.

The contributions of $N_{1,2}N_{1,2}$ are negligible due to the constraints
from neutrino oscillation data, LFV processes, and the DM relic density. This
requires the couplings $g_{1\alpha}$ and $g_{2\alpha}$ to be very small
compared to $g_{3\alpha}\sim\mathcal{O}(1)$ \cite{hna}. Moreover, the
contributions \{$\nu_{e}\bar{\nu}_{\tau}$, $\nu_{\tau}\bar{\nu}_{e}$,
$\nu_{\mu}\bar{\nu}_{\mu}$, $\nu_{\tau}\bar{\nu}_{\mu}$, $\nu_{\tau}\bar{\nu
}_{\tau}$\} are also negligible due the smallness of the couplings
$f_{\alpha\beta}$ in addition to the large value of $m_{S_{1}}$. Consequently,
the highly suppressed interactions mediated by $S_{1}$, makes the cross
section $\sigma(\mathcal{E}_{miss}\equiv\nu_{\mu}\bar{\nu}_{e})$ very close to
the background as illustrated in Fig. \ref{Svss}.

In order to maximize the signal detection, one has to choose a set of cuts
where the signal significance should be larger than 3 $\sigma$. The general
significance definition is defined by%
\begin{equation}
\mathcal{S}=N_{S}/\sqrt{N_{S}+N_{B}},\label{S}%
\end{equation}
where $N_{S}$ and $N_{B}$ are the signal and background events number,
respectively. Here $N_{S}$ is given by%
\begin{equation}
N_{S}=N_{EX}-N_{B}=L\times(\sigma^{EX}-\sigma^{BG}),
\end{equation}
where $N_{EX}$ is the expected events number, $L$ is the integrated
luminosity, and $\sigma^{EX}$ ($\sigma^{BG}$) is the expected (background)
cross section within the considered cuts.

One of the powerful characteristics of future $e^{+}e^{-}$ linear colliders,
such as the ILC, is the possibility of having the electron and/or positron
beams being polarized \cite{ILC4,ILC5}. This feature can be used to reduce the
background contribution which can result in a significant improvement of the
signal to background ratio. At the ILC, both electron and positron
polarizations are chosen to lie in the range \cite{ILC5}%
\begin{equation}
\left\vert P\left(  e^{-}\right)  \right\vert \leq0.8;~~\left\vert
P(e^{+})\right\vert \leq0.3,
\end{equation}
with $P\left(  f\right)  =(N_{f_{R}}-N_{f_{L}})/(N_{f_{R}}+N_{f_{L}})$; and
$N_{f_{R}}$ ($N_{f_{L}}$) is the number of right (left) handed fermions. For
CLIC, the positron polarization could reach up to $\left\vert P(e^{+}%
)\right\vert =0.6$; therefore one expects the background to be more suppressed
\cite{CLICP}\textit{.} Hence, by considering the electron (positron)
polarization $P\left(  e^{-}\right)  <0$ $(P(e^{+})>0)$, the excess in the
number of LH (RH) neutrino events gets enhanced.

We will carry out our simulation based on the benchmark (\ref{bA}), and use
the ILC run at different CM energy: $E_{CM}=$250 , 350, 500 \textrm{GeV} and 1
\textrm{TeV}, with unpolarized beams at first; then we consider the polarized
beams with $P\left(  e^{-},e^{+}\right)  =[-0.8,+0.3]$ and/or $P\left(
e^{-},e^{+}\right)  =[+0.8,-0.3]$. The details of our analysis is described
throughout the next section.

\section{Analysis and Discussion}

In this work, we used LanHEP \cite{LHP} and CalcHep \cite{CalcHEP} for the
simulation of our model, and generated the differential cross section with
respect to all the relevant kinematic variables. We found that the expected
cross section is barely larger than the background $\sigma^{EX}\gtrsim
\sigma^{BG}$, and the distributions have the same shape. We found out that the
useful kinematic variables, where the events number excess can be remarkable
are: the charged leptons energy ($E_{\ell}$), angular distributions
($\cos\theta_{\ell}$), the invariant mass ($M_{e,\mu}$), and the missing
invariant mass ($M_{miss}$). The latter variable can be reconstructed fairly
at any lepton collider since the full information about the initial state
momenta are provided. A summary of the considered cut sets is shown in
Table-\ref{T1}.

\begin{table}[h]%
\begin{tabular}
[c]{|c|c|}\hline\hline
$E_{CM}$ & Selection cuts\\\hline
$250$ & $%
\begin{array}
[c]{c}%
70~<E_{\ell}<110~,~70~<M_{e,\mu}<220~,~M_{miss}<120,\\
0.4621<\cos\theta_{e}<0.9640,~-0.9640<\cos\theta_{\mu}<-0.4621,
\end{array}
$\\\hline
$350$ & $%
\begin{array}
[c]{c}%
90~<E_{\ell}<165~,~100~<M_{e,\mu}<280~,~M_{miss}<200~,\\
0.4621<\cos\theta_{e}<0.9951,~-0.9866<\cos\theta_{\mu}<0,
\end{array}
$\\\hline
$500$ & $%
\begin{array}
[c]{c}%
120~<E_{\ell}<240~,~300~<M_{e,\mu}<480~,~M_{miss}<300~,\\
0.4621<\cos\theta_{e}<0.9951,~-0.9951<\cos\theta_{\mu}<0,
\end{array}
$\\\hline
$1000$ & $%
\begin{array}
[c]{c}%
E_{\ell}<70~,~M_{e,\mu}<140~,~M_{miss}>750~,\\
0.0997<\cos\theta_{e}<0.6640,~-0.6640<\cos\theta_{\mu}<-0.0997.
\end{array}
$\\\hline\hline
\end{tabular}
\caption{Relevant cuts for the process $e^{+}e^{-}\rightarrow E_{miss}%
+e^{-}\mu^{+}$ at different CM energies. Here $E_{\ell}$ and $\theta_{\ell}$
are the charged lepton energy in emission angles, $M_{e,\mu}$ is the
electron-muon invariant mass and $M_{miss}$ is the missing invariant mass. All
masses and energies are given in \textrm{GeV}.}%
\label{T1}%
\end{table}

From the angular cuts at different CM energies in Table-\ref{T1}, the charged
leptons from the process $e^{+}+e^{-}\rightarrow e^{-}\mu^{+}+E_{miss}$ could
be emitted in wide angle ranges, while in a similar model studied in
\cite{kan}, the outgoing leptons ($e^{-}$ and $\mu^{+}$) are emitted almost
collinearly. This is due to the fact that in our model the process proceeds
via both $t$- and $s$-channel whereas in \cite{kan} it is via only the $t$-channel.

By imposing the cuts in Table-\ref{T1} at each CM energy, we obtain both the
expected signal and the background cross sections. This gives us an idea about
the required luminosity from the significance values as presented in
Table-\ref{T2}.

\begin{table}[h]%
\begin{tabular}
[c]{|c|c|c|c|c|c|}\hline\hline
$E_{CM}$ & $\sigma^{BG}$ & $\sigma^{EX}$ & $\left(  \sigma^{EX}-\sigma
^{BG}\right)  /\sigma^{BG}$ & $\mathcal{S}_{100}$ & $\mathcal{S}_{500}%
$\\\hline
250 & $6.5919\times10^{-2}$ & $6.7402\times10^{-2}$ & $2.2497\times10^{-2}$ &
$1.8064$ & $4.0391$\\\hline
350 & $5.8882\times10^{-2}$ & $6.0158\times10^{-2}$ & $2.2723\times10^{-2}$ &
$\allowbreak1.6451$ & $3.6787$\\\hline
500 & $5.6560\times10^{-2}$ & $5.7630\times10^{-2}$ & $1.8918\times10^{-2}$ &
$1.4095$ & $3.1517$\\\hline
1000 & $1.9217\times10^{-5}$ & $4.6976\times10^{-4}$ & $23.445$ & $6.5735$ &
$14.699$\\\hline\hline
\end{tabular}
\caption{The cross sections of the total expected signals and the background
estimated for the considered energies within the cuts given in Table-\ref{T1};
and the significance $\mathcal{S}_{100}$ and $\mathcal{S}_{500}$ that
correspond to the two integrated luminosity values L=100, 500 $fb^{-1}$,
respectively. All energies are given in \textrm{GeV} and cross sections in
\textrm{pb}.}%
\label{T2}%
\end{table}

One has to mention that for $E_{CM}=$250, 350 and 500 \textrm{GeV}, the
corresponding required luminosity to detect the signal should be higher than
the values reported in \cite{ILC3}, in contrast to the case where $E_{CM}=$1
\textrm{TeV}. As can be seen from the cuts on the charged leptons energy and
the missing invariant mass in Table-\ref{T1} in addition to the cross sections
in Table-\ref{T2}, clearly the events number excess at $E_{CM}=1$ \textrm{TeV}
has a different source compared to the other CM energies. Here, the missing
energy is mainly RH Majorana neutrinos, due to the following reasons: (1) The
missing invariant mass is large because $N_{3}$ is very massive, (2) the
existence of heavy RH neutrinos in the final state leads to a small phase
space for the daughter particles including the charged leptons, and (3) the
expected cross section is dominated by the subprocess $\mathcal{E}%
_{miss}\equiv N_{3}N_{3}$, whereas the $\mathcal{E}_{miss}\equiv\nu_{\mu}%
\bar{\nu}_{e}$ contribution is comparable to the background.

\begin{figure}[t]
\begin{centering}
\includegraphics[width=5.8cm,height=4.2cm]{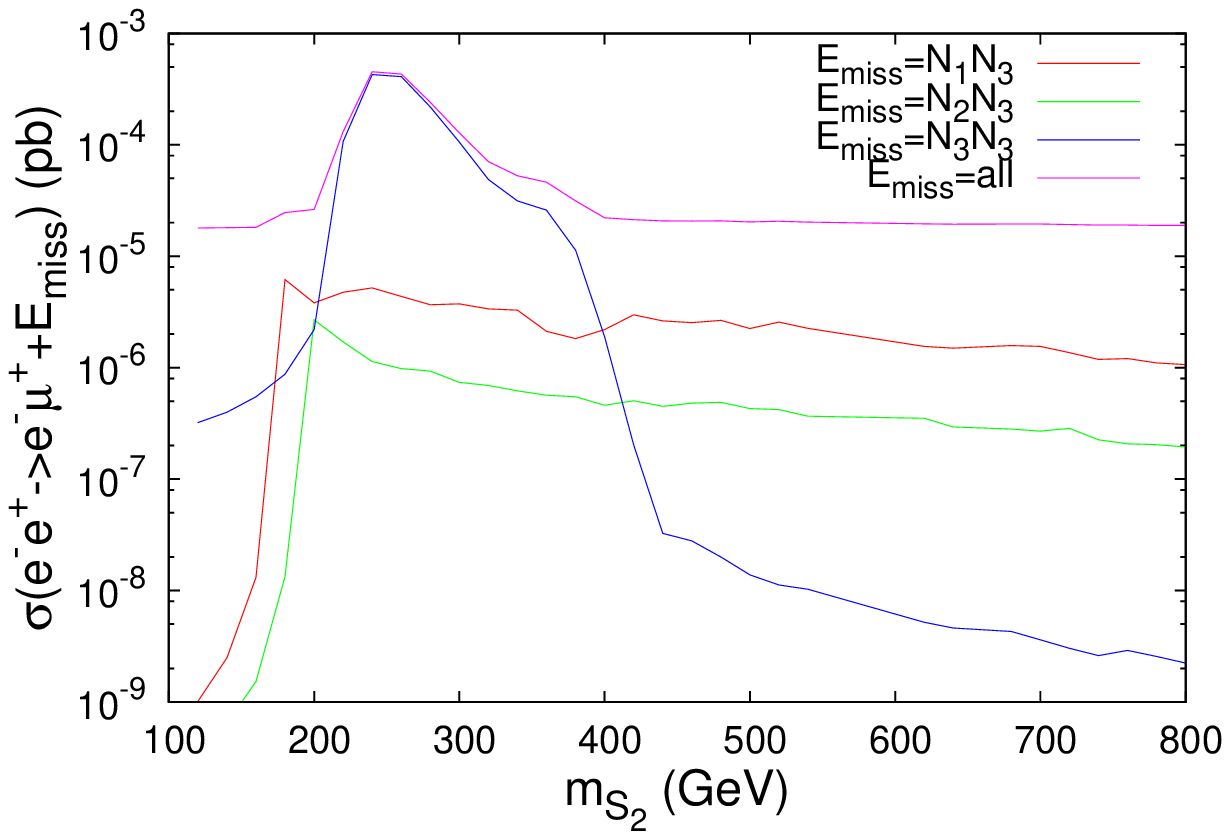}\includegraphics[width=5.8cm,height=4.2cm]{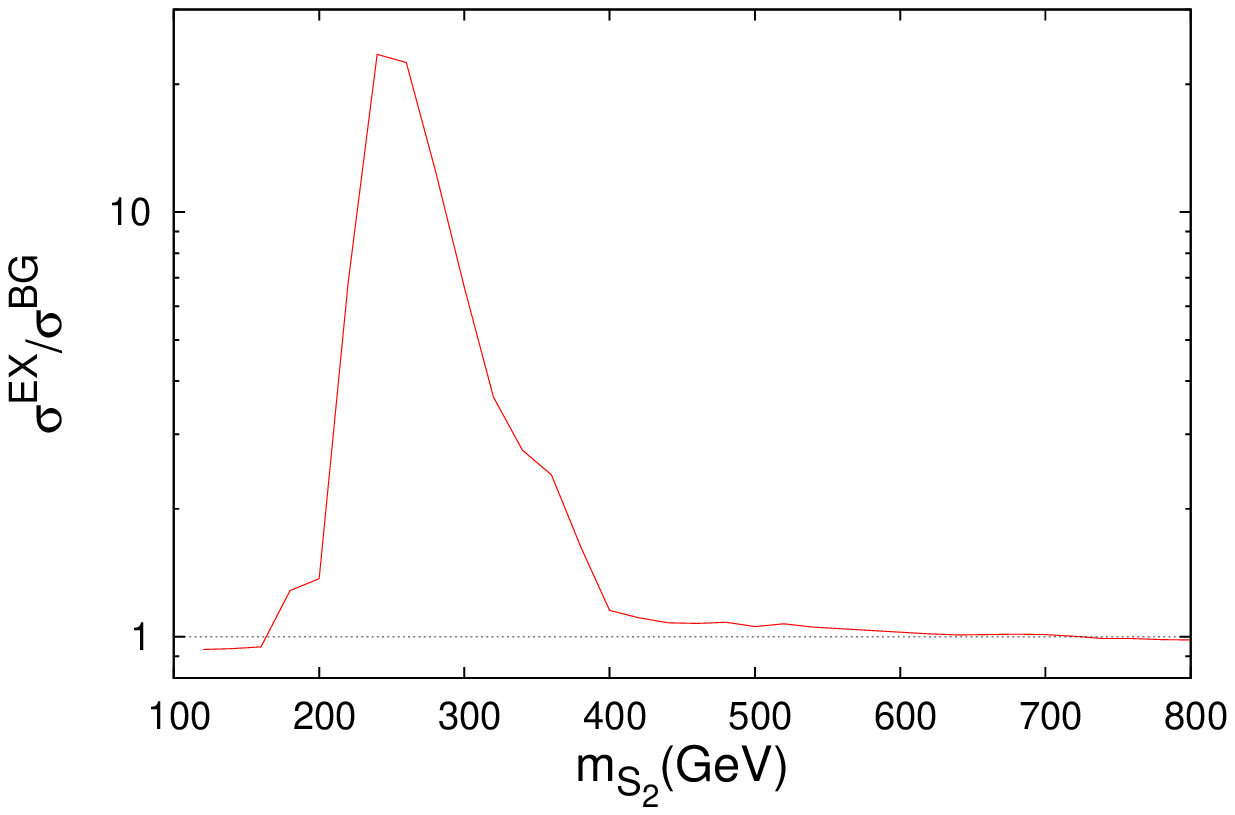}\includegraphics[width=5.8cm,height=4.2cm]{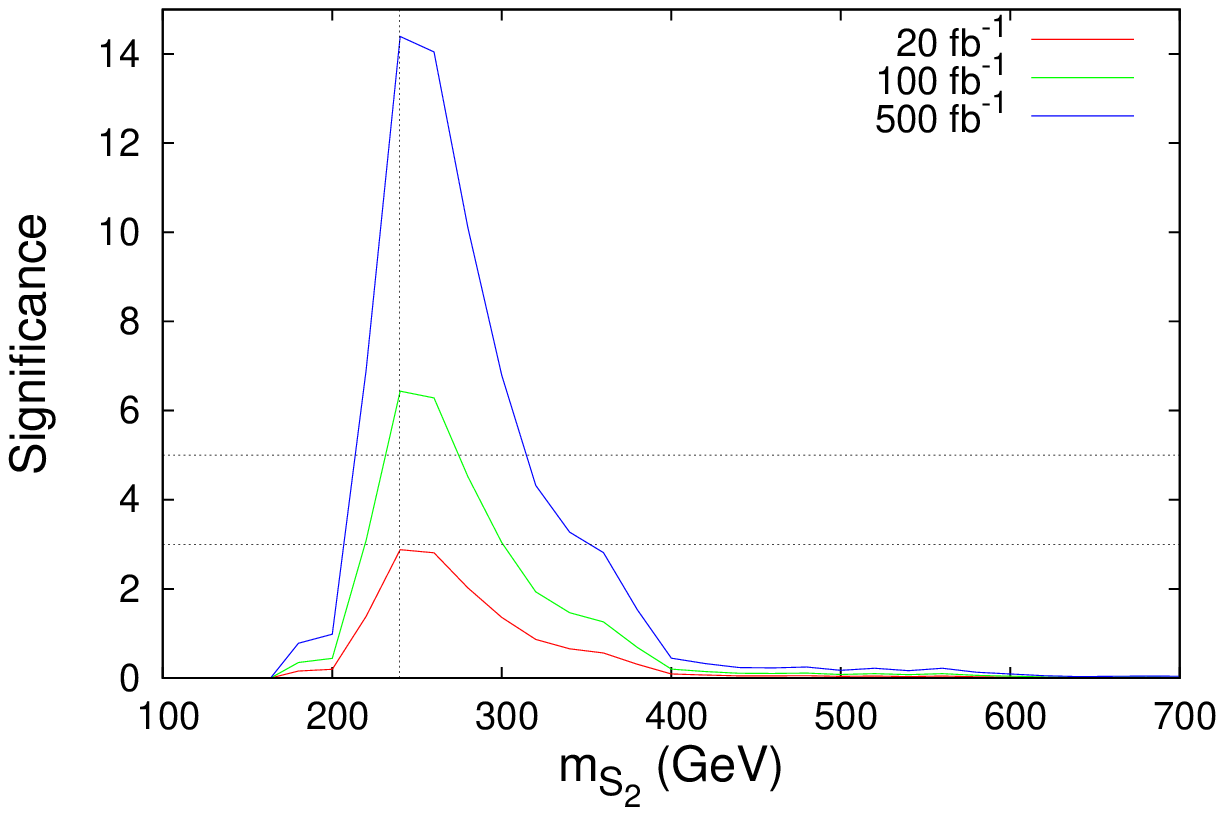}
\par\end{centering}
\caption{These plot are obtained for $E_{CM}$=1 \textrm{TeV} within the cuts
given in Table-\ref{T1}. Left panel: The cross section of different subprocess
as a function of the charged scalar masses $m_{S_{2}}$. Middle panel: The
expected cross section $\sigma^{EX}$ scaled by the background. Right panel:
The significance ($\mathcal{S}$) versus $m_{S_{2}}$ for different integrated
luminosity values, where the two dashed horizontal lines represent the values
$\mathcal{S}=3$ and $\mathcal{S}=5$, respectively; and the vertical one
represents the physical value of the charged scalar mass $m_{S_{2}}$ (given in
(\ref{bA})).}%
\label{ms2}%
\end{figure}

Note that there are 34 Feynman diagrams that contribute to the amplitude of
the subprocess $e^{-}e^{+}\rightarrow e^{-}\mu^{+}+N_{3}N_{3}$, all of them
mediated by the charged scalar $S_{2}$.\textbf{\ }Then, in Fig. \ref{ms2}, we
illustrate the cross sections of the different $\mathcal{E}_{miss}\equiv
N_{i}N_{3}$ subprocesses, and the corresponding significance versus the
charged scalar mass $m_{S_{2}}$.

Moreover in Fig. \ref{ms2}, the cross section $\sigma(\mathcal{E}_{miss}\equiv
N_{3}N_{3})$ gets enhanced for the charged scalar mass near the resonance
value $m_{S_{2}}\sim$240 \textrm{GeV}, whereas it is negligible elsewhere. It
is interesting to notice that the $m_{S_{2}}$ value in our benchmark
(\ref{bA}) corresponds exactly with the resonance. Here again by looking at
the charged scalar masses $m_{S_{2}}<164$ \textrm{GeV} and $m_{S_{2}}>724$
\textrm{GeV}, the cross section $\sigma^{EX}$ is expected to be smaller than
the background, and therefore there is no significant events number excess
that can be observed in this case.

Now, we extend our study by considering polarized beams in order to increase
the signal to the background. Hence, we consider $P\left(  e^{-},e^{+}\right)
=[-0.8,+0.3]$ for $E_{CM}=$250, 350, and 500 \textrm{GeV} and $P\left(
e^{-},e^{+}\right)  =[+0.8,-0.3]$ for $E_{CM}=1$ \textrm{TeV}. Keeping the
same defined cuts listed in Table-\ref{T1}, the estimated cross section and
significance values presented in Table-\ref{T2} get modified, as summarized in
Table-\ref{T3}.

\begin{table}[ptb]%
\begin{tabular}
[c]{|c|c|c|c|c|c|c|}\hline\hline
$E_{CM}$ & P($e^{-},e^{+}$) & $\sigma^{BG}$ & $\sigma^{EX}$ & $\left(
\sigma^{EX}-\sigma^{BG}\right)  /\sigma^{BG}$ & $\mathcal{S}_{100}$ &
$\mathcal{S}_{500}$\\\hline
250 & -0.8, +0.3 & $0.15399$ & $0.15910$ & $3.3184\times10^{-2}$ & $4.0512$ &
$9.0588$\\\hline
350 & -0.8, +0.3 & $0.13640$ & $0.13997$ & $2.6173\times10^{-2}$ & $3.0175$ &
$6.7474$\\\hline
500 & -0.8, +0.3 & $0.13100$ & $0.13450$ & $2.6718\times10^{-2}$ & $3.0179$ &
$6.7483$\\\hline
1000 & +0.8, -0.3 & $2.0708\times10^{-6}$ & $7.2710\times10^{-4}$ & $350.12$ &
$8.5027$ & $19.013$\\\hline\hline
\end{tabular}
\caption{The cross sections for the total expected signals and the background
estimated for the considered energies within the cuts given in Table-\ref{T1},
and the significance $\mathcal{S}_{100}$ and $\mathcal{S}_{500}$ that
corresponds to the two integrated luminosity values $L$=100, 500 $fb^{-1}$,
respectively. All energies are given in \textrm{GeV} and cross sections in
\textrm{pb}.}%
\label{T3}%
\end{table}

\begin{figure}[t]
\begin{centering}
\includegraphics[width=8cm,height=5.5cm]{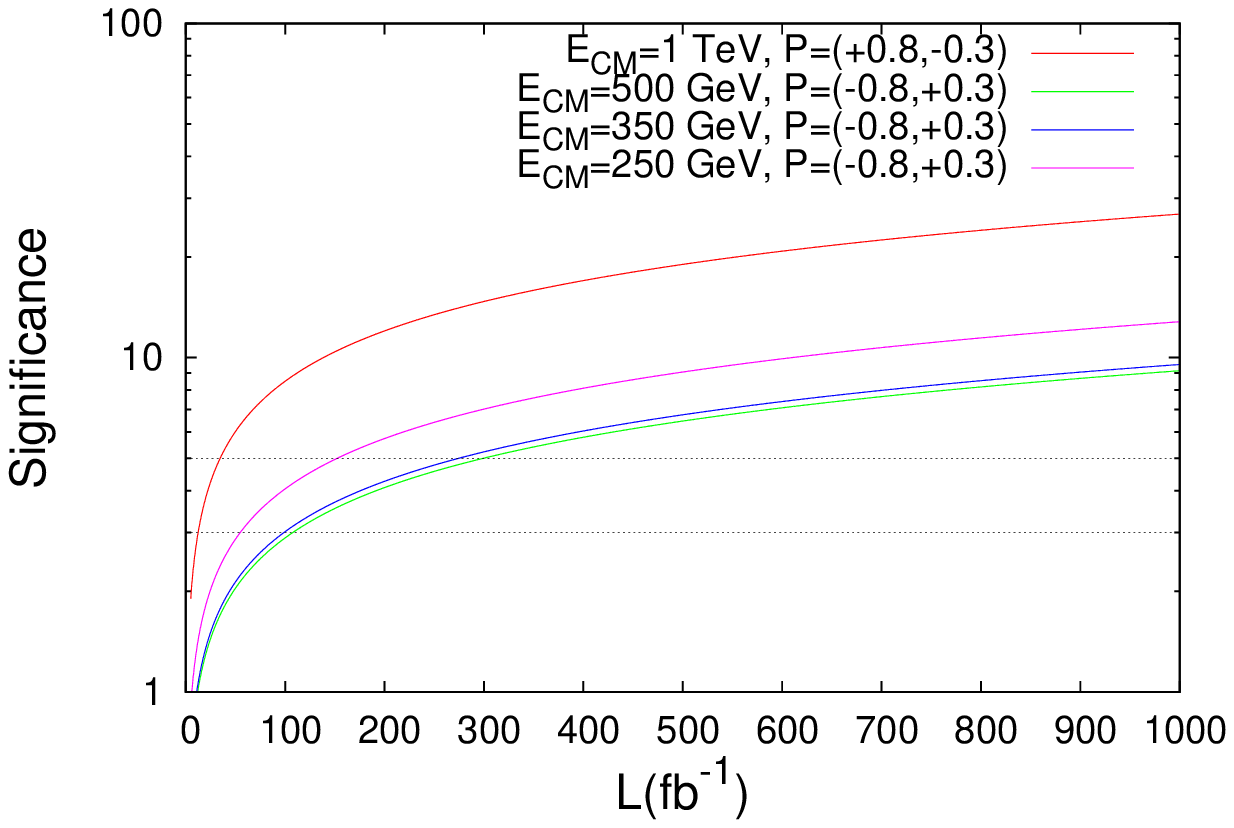}\includegraphics[width=8cm,height=5.5cm]{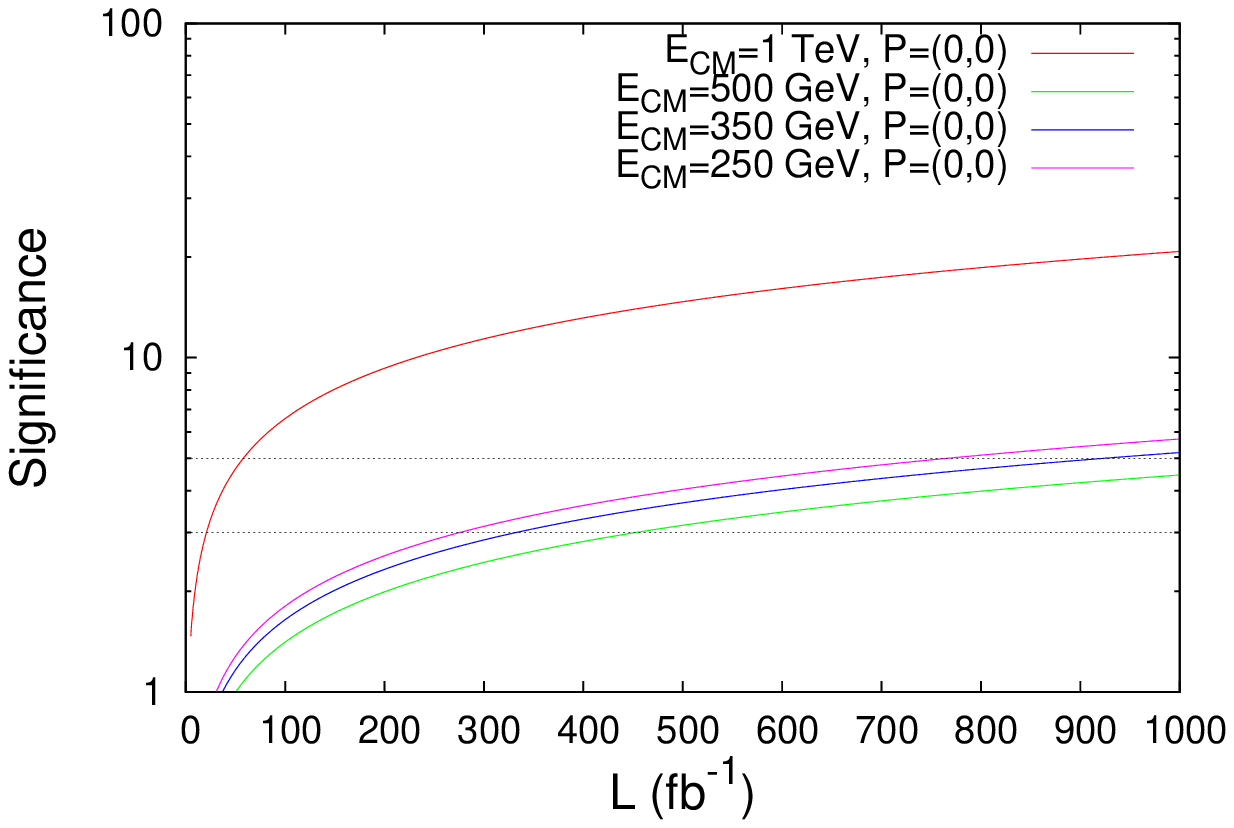}
\par\end{centering}
\caption{The significance versus luminosity at different CM energies within
the cuts defined in Table-\ref{T1}; with (left) and without (right) polarized
beams. The two horizontal dashed lines represent $\mathcal{S}=3$ and
$\mathcal{S}=5$, respectively.}%
\label{SvsL}%
\end{figure}

From Table-\ref{T3}, after using the polarization, the expected cross section
gets enhanced by about 150\% for $E_{CM}=$250, 350, 500 \textrm{GeV}, and by
about 50\% for $E_{CM}=1$ \textrm{TeV}. This makes the signal easy to detect
for all the considered CM energies. As a summary, we give in Table-\ref{T4}
the expected events number excess for each CM energy with and without
polarized beams. In Fig. \ref{SvsL}, we show the dependance of the
significance on the accumulated luminosity with and without polarized beams
for the considered CM energies, within the cuts given in Table-\ref{T1}. Thus,
we see that by having a polarized beam, the signal can be observed even with
relatively low integrated luminosity. For example, at $E_{CM}=250$
\textrm{GeV}, the 5 $\sigma$ required luminosity is 150 $fb^{-1}$ for the
polarized beam as compared to 700 $fb^{-1}$ without polarization.

\begin{table}[ptb]%
\begin{tabular}
[c]{|c|c|c|c|c|c|}\hline\hline
$E_{CM}$ $($\textrm{GeV}$)$ & $L$ $(fb^{-1})$ & $P(e^{-},e^{+})$ & $N_{B}$ &
$N_{EX}$ & $N_{S}$\\\hline
$250$ & $250$ & $0,0$ & $16480$ & $16851$ & $371$\\\cline{3-6}
&  & $-0.8,+03$ & $38498$ & $39775$ & $1277$\\\hline
$350$ & $350$ & $0,0$ & $20609$ & $21055$ & $446$\\\cline{3-6}
&  & $-0.8,+03$ & $47740$ & $48990$ & $1250$\\\hline
$500$ & $500$ & $0,0$ & $28280$ & $28815$ & $535$\\\cline{3-6}
&  & $-0.8,+03$ & $65500$ & $67250$ & $1750$\\\hline
$1000$ & $1000$ & $0,0$ & $19.217$ & $469.76$ & $450.54$\\\cline{3-6}
&  & $+0.8,-03$ & $2.07$ & $727.10$ & $725.03$\\\hline\hline
\end{tabular}
\caption{The expected ($N_{EX}$) and background ($N_{B}$) number of events for
different CM energy values with/without polarized beams within the cuts given
in Table-\ref{T1}.}%
\label{T4}%
\end{table}

\section{Conclusions}

In this work, we have studied the detectability of a radiative model for
neutrino masses at the future $e^{-}e^{+}$ linear colliders. At different CM
energies $E_{CM}=$250, 350, 500 \textrm{GeV} and 1 \textrm{TeV}, we studied
the process $e^{-}e^{+}\rightarrow e^{-}\mu^{+}+E_{miss}$. We found that for
the CM energies 250, 350 and 500 \textrm{GeV}, the missing energy is mainly
light LH neutrinos, while at $E_{CM}=1$ \textrm{TeV}, the RH neutrinos
contribution is dominant.

We have shown for the CM energy $E_{CM}=1$ \textrm{TeV}, the signal is
sensitive to the mass of the charged scalar $m_{S_{2}}$; and it is more
significant near a resonant value $m_{S_{2}}\sim240$ \textrm{GeV}, which
corresponds to the value chosen in our benchmark.

We found that the signal cannot be seen at $E_{CM}=$250, 350, 500 \textrm{GeV}
in contrast the case of $E_{CM}=1$ \textrm{TeV}. After using polarized beams,
the signal gets enhanced and can be observed for all CM energies. Furthermore,
when considering polarization, the signal can be detected with smaller
integrated luminosity as compared to unpolarized beam case.

\begin{acknowledgements}
We thank the ICTP and CERN for their hospitality where a large part
of this work has been carried out. We thank S. Kanemura for useful
discussions during early stages of this work, and K. Yagyu for
reading the manuscript. Special thanks to A. Djouadi for the careful
reading and the very useful discussions. The work of A. A. is
supported by the Algerian Ministry of Higher Education and
Scientific Research under the PNR '\textit{Particle Physics /
Cosmology: the interface}'; and the CNEPRU Project No. D01720130042.
\end{acknowledgements}

\end{document}